\documentclass[aps,eqsecnum,amsmath,onecolumn,groupedaddress,superscriptaddress,notitlepage,nofootinbib]{revtex4-1}
\usepackage[utf8]{inputenc}
\usepackage{indentfirst}
\usepackage{amssymb}
\usepackage{amsmath,latexsym}
\usepackage{graphicx}
\usepackage{float}
\usepackage[margin=1.0in]{geometry}
\usepackage[mathscr]{euscript}
\usepackage{color}
\usepackage{xcolor}
\usepackage{soul}
\usepackage{slashed}
\graphicspath{{images/}}
\usepackage[mathscr]{euscript}
\DeclareMathOperator{\Tr}{Tr}

\newcommand{\beq}{\begin{equation}}
\newcommand{\eeq}{\end{equation}}
\newcommand{\bea}{\begin{eqnarray}}
\newcommand{\eea}{\end{eqnarray}}

\def\beqs#1\eeqs{\beq\begin{split} #1 \end{split}\eeq}

\def\spl#1{\begin{split} #1 \end{split}}

\def\pd#1#2{\frac{\partial #1}{\partial #2}}

\def\Im {\mathop{\hbox{Im}}}
\def\Tr {\mathop{\hbox{Tr}}}

\def\av#1{ \left\langle #1 \right\rangle }

\long\def\comment#1{}

\usepackage[colorlinks=true,backref=false, linktocpage=true,
citecolor=blue,urlcolor=blue,linkcolor=blue,pdfpagemode=UseOutlines]{hyperref}

\hypersetup{%
  bookmarksnumbered=true,
  pdftitle = {},
  pdfsubject = {},
  pdfauthor = {},
  pdfkeywords = {}
}

\usepackage{microtype}

\newcommand{\eq}[1]{Eq.~(\ref{#1})}
\newcommand{\fig}[1]{Fig.~\ref{#1}}

\begin{document}
\title{Spin Polarized Non-Relativistic Fermions in 1+1 Dimensions}

\author{Andrei Alexandru}
\email{aalexan@gwu.edu }
\affiliation{Department of Physics, The George Washington University,
Washington, DC 20052}
\affiliation{Department of Physics,
University of Maryland, College Park, MD 20742}
\author{Paulo F. Bedaque}
\email{bedaque@umd.edu}
\author{Neill C. Warrington}
\email{ncwarrin@umd.edu}
\affiliation{Department of Physics,
University of Maryland, College Park, MD 20742}

\date{\today}

\begin{abstract}
We study by Monte Carlo methods the thermodynamics of a spin polarized gas of non-relativistic fermions in 1+1 dimensions.  The main result of this work is that our action suffers no significant sign problem for any spin polarization in the region relevant for dilute degenerate fermi gases. This lack of sign problem allows us to study attractive spin polarized fermions non-perturbatively at spin polarizations not previously explored. For some parameters values we
verify results previously obtained by methods which include an uncontrolled step like complex Langevin and/or analytical continuation from imaginary chemical potential.
For others, larger values of the polarization, we deviate from these previous results.
\end{abstract}


\maketitle

\section{Introduction}

Strongly coupled dilute fermi gases are rich systems which stand at the crossroads of various fields of physics. In the realm of nuclear physics, it is believed that the properties of dilute fermi gases with large scattering lengths are related to the properties of neutron rich matter at the edge of a neutron star~\cite{2003AIPC..690..184C}. In the context of core collapse supernovae, the dynamic properties of dilute neutron gases must be analyzed to understand the propagation of energy which drive the explosion~\cite{Burrows:2004vq,Bedaque:2018wns}. In the atomic physics community, dilute fermi gases at unitarity have been studied to in the context of the BEC-BCS crossover~\cite{Zwerger:2012}. Spin polarizing these gases adds a layer of richness to observable phenomena. For instance, unequal density for the two spin states may either lead to an instability of Cooper pairing \cite{Bedaque:2003hi} or the formation of the exotic superconducting phase predicted by Larkin, Ovchinnikov, Fulde, and Ferrell~\cite{PhysRev.135.A550,Larkin:1965}, characterized by formation of Cooper pairs which spontaneously break translation invariance. High magnetic fields in neutron stars can destroy $S$-wave pairing in neutron matter through spin polarization, which has observable consequences on spectroscopic data gathered from rapidly rotating neutron stars~\cite{PhysRevC.93.015802}. 

All these systems are strongly coupled and to investigate their properties requires non-perturbative methods. One general approach to analyzing strongly coupled systems with controllable systematic uncertainties are \emph{ab initio} Monte Carlo calculations. This approach runs into difficulties for spin imbalanced non-relativistic systems due to the \emph{sign problem} which causes statistical uncertainties to grow exponentially as the thermodynamic and zero temperature limits are taken. The source of the sign problem in fermionic systems are sign oscillations of the fermion determinant. 

A few suggestions for dealing with the sign problem in spin polarized non-relativistic systems have been put forward in the past. One  is to compute observables at imaginary chemical potential, which renders the fermion determinant positive definite, and hence eliminates the sign problem and analytically continue them to real chemical potentials. 
The process of analytic continuation from restricted, noise data is necessarily done with the help of an ansatz which introduces an uncontrolled source of error. Still, mean field results suggest that interesting features of the phase diagram lie within the region where extrapolation works~\cite{Braun:2012ww}. For the one dimensional case, this avenue has been tested numerically~\cite{PhysRevA.92.063609}. Another approach is the Complex Langevin method where the configuration space is augmented by complexifying the field variables and stochastic process is used to sample configurations in this complex space~\cite{Rammelmuller:2017myk,Loheac:2017uuj}. Complex Langevin has also been used to analyze non-relativistic fermions with a mass imbalance, which is a system that also suffers from the sign problem \cite{PhysRevD.96.094506}.
The potential problem of complex Langevin calculations is that it may converge to the wrong result and/or lead to runaway solutions~\cite{Aarts:2010aq,PhysRevD.88.116007,Aarts:2011ax}. The problem of runaway solutions was cleverly taken care, in the one-dimensional non-relativistic case, by modifying the action with an extra term whose strength is taken to zero at the end \cite{Drut:2017fsv,PhysRevD.95.094502}. It would be important to verify and validate this method by independent means.

In the present work we show that,  the sign problem is very mild in the parameter region investigated using other methods~\cite{PhysRevA.92.063609,Rammelmuller:2017myk} and direct investigations of the polarized systems can be performed. We study the simple case of fermions interacting through an attractive s-wave interaction in $1+1$ dimensions, with an eye toward extending our work to atomic and nuclear systems 2+1 and 3+1 dimensions. Preliminary analysis indicate that the sign problem is also very mild in higher spatial dimensions as well.

This paper is organized as follows: In Section~\ref{the-model} we discuss the model and its
lattice discretization. In Section~\ref{method} we describe the details of our Monte Carlo simulations. In Section~\ref{results} we present the results of our simulations. Lastly, in Section~\ref{conclusions} we summarize our findings and discusses future prospects.

\section{The Model}\label{the-model}

\def\up{\uparrow}
\def\down{\downarrow}
We study the thermodynamics of a dilute system of non-relativistic spin-1/2 fermions 
interacting via a contact interaction. The Hamiltonian of the system is 
\beq
H =  \int dx\, \left[ \sum_{\sigma=\up,\down} \psi_\sigma^\dagger(x) \left(-\frac{\nabla^2}2 \right)
 \psi_\sigma(x) - G\ n_\up(x) n_\down(x) \right] \,,
\eeq
where $\sigma$ runs over the fermion polarizations and 
$n_\sigma(x)=\psi_\sigma^\dagger(x)\psi_\sigma(x)$ is the fermion density.
We use non-relativistic natural units $\hbar = M = k_B = 1$ with 
$M$ being the  mass of the fermion. We take the interaction to be attractive so $G>0$. 
In one spatial dimension, the s-wave coupling is cutoff independent since fermion-fermion loops are ultraviolet finite. $G$ is related to the binding energy of the unique two-particle bound state by $E= -{G^2}/{4}$ (in the continuum, infinite volume limit).

The grand canonical partition 
function for the system can be written using the path integral formulation:
\beq
Z = \Tr e^{-\beta(H-\sum_\sigma \mu_\sigma N_\sigma)} =
\int{D\psi D\psi^{\dagger}e^{-S[\psi,\psi^{\dagger}]}}\,,
\eeq
with $N_\sigma=\int dx\, n_\sigma(x)$. In the path integral formulation the integrand
involves the Euclidean action
\beq
S = \int_{0}^{\beta}\int{d\tau dx\Bigg[\sum_\sigma (\psi_{\sigma}^{\dagger}\partial_{\tau}\psi_{\sigma} - \psi_{\sigma}^{\dagger}\frac{\nabla^2}{2}\psi_{\sigma}- \mu_\sigma\psi_{\sigma}^{\dagger}\psi_{\sigma}) - G(\psi_\up^{\dagger}\psi_\up )(\psi_\down^{\dagger}\psi_\down )\Bigg]} \,,
\label{eq:continuum-action}
\eeq 
The chemical potentials are defined as $\mu_\up=\mu+h$, $\mu_\down=\mu-h$. $\mu$ controls the density of particles while the spin chemical potential $h$ controls the spin polarization. 
Note that in the path integral the fields $\psi$ are Grassman fields, but we use the same
notation as the operator fields appearing in the Hamiltonian to keep the notation simple.

For numerical simulations the quartic fermion term associated with the contact interaction
is transformed into a quadratic term via a Hubbard-Stratonovich transformation~\cite{Hubbard:1959ub}. The discretized action we start with is
\beqs
S =& \frac{1}{\hat g}\sum_{x,t}{\Big[\cosh(A_{x,t})-1\Big]}+\sum_{\sigma;x,t} \Bigg[{\hat \psi^{\dagger}_{\sigma;x,t}\hat \psi_{\sigma;x,t}}- {\hat \psi^{\dagger}_{\sigma;x,t+1}\text{e}^{A_{x,t}+ \hat \mu_\sigma}\hat \psi_{\sigma;x,t}}\\
&+ \frac{\hat \gamma}{2}{ {(\hat \psi^{\dagger}_{\sigma;x+1,t}-\hat \psi^{\dagger}_{\sigma;x,t})(\hat \psi_{\sigma;x+1,t}-\hat \psi_{\sigma;x,t})}}\Bigg]
\label{eq:lattice-action-dimless}
\eeqs 
where $A_{xt}$ is a real, bosonic auxiliary field and $\hat\mu_\sigma = \hat\mu \pm \hat h$.
All hatted quantities denote lattice quantities. The discretized action converges to the 
right continuum limit when
\beq
\spl{
\mu \Delta t &= \frac{f_1}{f_0}\text{e}^{\hat \mu}\cosh\hat h-1\\
h \Delta t & = \frac{f_1}{f_0}\text{e}^{\hat \mu}\sinh \hat h\\
\frac{G \Delta t}{\Delta x}&= \left(\frac{f_2}{f_0}-\frac{f^2_1}{f^2_0} \right)e^{2\hat \mu} \\
\frac{\Delta t}{\Delta x^2} &= \hat \gamma  }\,,
\qquad\text{or inversely}\qquad
\spl{
&\hat\gamma = \Delta t/\Delta x^2 \\
&\zeta = (\mu \Delta t+1)^2-(h\Delta t)^2 \\
&\hat h = {\rm arcsinh\,}( h\Delta t/\sqrt\zeta)\\
&\frac{f_1^2}{f_0 f_2} = \frac\zeta{\zeta+G\Delta t/\Delta x} (\text{solve for $\hat g$})\\
&\hat\mu = \log (\sqrt\zeta f_0/f_1)
}\, ,
\label{eq:phys-coups}
\eeq
where
\beq
f_\alpha(\hat g) \equiv \int_{-\infty}^\infty dA \exp\left(-\frac{\cosh A-1}{\hat g}+\alpha A\right)  .
\eeq The matching
conditions above can be derived by integrating out the auxiliary fields using the relation
\beqs
&\int_{-\infty}^\infty dA \exp\left(-\frac{\cosh A-1}{\hat g}\right) \, \exp[- \hat\psi^\dagger e^{A+\mu} \hat\psi]
=\int_{-\infty}^\infty dA \exp\left(-\frac{\cosh A-1}{\hat g}\right)\, [ 1 - \hat\psi^\dagger e^{A+\mu} \hat\psi
+\frac{1}{2} (\hat\psi^\dagger e^{A+\mu} \hat\psi)^2 ] \\
&\quad= f_0 - f_1 \hat\psi^\dagger e^{\mu} \hat\psi +\frac{1}{2} f_2 (\hat\psi^\dagger e^{\mu} \hat\psi)^2
= f_0 \exp \left[\frac{f_1}{f_0}\hat\psi^\dagger e^{\mu} \hat\psi + \left(\frac{f_2}{f_0}-\frac{f_1^2}{f_0^2} \right) \frac{1}{2}(\hat\psi^\dagger e^{\mu} \hat\psi)^2 \right] \,,
\eeqs
and then take the continuum limit $\Delta t,\Delta x\to0$.  The continuum time and space limits can be taken separately. If we take the  continuum time limit first
we get the partition function for a discretized fermionic system with on-site interactions,
an {\em attractive} version of the Hubbard model, with Hamiltonian
\beq
H_\text{disc} = \sum_x \left\{ \sum_\sigma \left[\frac1{2\Delta x^2} (\hat\psi_{\sigma;x+1}^\dagger-
\hat\psi_{\sigma;x}^\dagger)(\hat\psi_{\sigma;x+1}-
\hat\psi_{\sigma;x}) - \mu_\sigma \hat n_{\sigma;x} \right]-\frac G{\Delta x} \hat n_{\up,x}
\hat n_{\down,x} \right\} \,.
\eeq
Note that above $\hat\psi$'s are operators normalized to satisfy the canonical commutation relations 
$\{\hat\psi_{\sigma;x}^\dagger, \hat\psi_{\sigma';x'}\} = \delta_{\sigma\sigma'}\delta_{xx'}$.
The discretized partition function converges linearly, as ${\cal O}(\Delta t)$, 
to the time continuum limit. We will show this explicitly and discuss how to improve the $\Delta\rightarrow 0$ convergence later. When taking the spatial 
continuum limit, the discretized Hamiltonian above contain errors of order $\Delta x^2$.
In any case, the convergence rate does not change if we replace the matching conditions
above with their first order approximation in the $\Delta t\to 0$ limit:
\beq
\hat\gamma=\frac{\Delta t}{\Delta x^2}\,,\quad
\hat h = h\Delta t \,, \quad
\hat\mu= \mu\Delta t - \frac{G\Delta t}{2\Delta x}\,,\quad\text{and}\quad
\hat g = \frac{G\Delta t}{\Delta x} \,.
\eeq
We will use these matching conditions for our simulations.

 
\section{Numerical method}\label{method}
 
The fermionic part of the action is 
quadratic in the fields and we can compute analytically the integral over the fermionic variables. 
The partition function resulting is then
\beq
Z = \int \prod_{x,t} dA_{x,t}\, e^{-S_g(A)} \det D(A) \,,
\eeq
where $S_g(A) = \sum_{x,t}(\cosh A_{x,t}-1)/\hat g$ and $D(A)$ is the fermionic matrix appearing
in Eq.~\ref{eq:lattice-action-dimless}. The matrix is block diagonal in spin space, that
is $D_{\sigma xt,\sigma'x't'} = \delta_{\sigma\sigma'} (D_\sigma)_{xt,x't'}$ with
\beqs
&(D_\sigma)_{xt,x't'}=B_{xx'}\delta_{tt'} -
e^{\hat\mu_\sigma} C_{xx'}(A_t) \delta_{t,t'+1} \,, \\
&B_{xx'} = (1+ \hat \gamma) \delta_{xx'} - \frac{\hat \gamma}{2}(\delta_{x,x'+1}+\delta_{x,x'-1}) \,,\qquad
C(A_t)_{xx'} = \delta_{x,x'}\text{e}^{A_{xt}}\,.
\eeqs
Above, $A_t$ indicates the slice of the lattice field $A$  at time $t$. Note that the
dependence on spin appears only as a factor multiplying the temporal hopping matrix $C$. Since
the determinant is diagonal in spin space we have $\det D(A) = \det D_\up(A) \det D_\down(A)$.
When $\hat h=0$ then $\mu_\up=\mu_\down$ and $\det D(A) = \det D_\up(A)^2>0$, a positive 
quantity since $D_\sigma$'s are real matrices. For the spin polarized case, 
$\hat h \neq 0$, the determinant is no longer always
positive raising the possibility of a sign problem. The sampling will then be done using
the positive probability distribution
\beq
p_\text{pq}(A) = \frac{1}{Z}e^{-S_g(A)} |\det D(A)|\,.
\label{eq:prob-dist}
\eeq
The sign of the determinant is then included in the measured observable
\beq
\av{ \mathcal{O}} = \frac{\av{\mathcal{O}\,\text{sign}\, D}_\text{pq}}
{\av{ \text{sign}\,D}_\text{pq}} \,.
\eeq 
Above $\av{\cdot}_\text{pq}$ indicates averages with respect to the {\em phase quenched} 
probability distribution $p_\text{pq}(A)$.
An important observation in this paper is that for all simulations described here {\em we did not
encounter a negative determinant}. This is not to say that there are no $A$ configurations
that lead to negative determinants, but that for the parameter region used in this study
the probability of such configurations is extremely small.
 
By for the most expensive step of the calculation is the computation of the fermion determinant, so we have optimized its computation. 
The fermion matrix, when the entries corresponding to a time-slice are grouped together, has
the following structure
\beq
D_{\sigma} = 
\begin{bmatrix}
B & 0 & \dots & e^{\hat\mu_\sigma} C(A_{N_t -1})\\
-e^{\hat\mu_\sigma} C(A_{0}) & B & \dots & 0\\
\vdots & \ddots & \ddots & \vdots\\
0 & \dots & -e^{\hat\mu_\sigma} C(A_{N_t -2}) & B
\end{bmatrix} \,.
\label{eq:block}
\eeq 
Using the sparsity of the temporal blocks, the calculation of the determinant $\det D_\sigma$
can be reduced to a smaller problem using Schur's complement methods, similar to the procedure
used for Wilson fermions in lattice QCD~\cite{Alexandru:2010yb} (note that a similar expression
was derived earlier for the Hubbard model using operator 
methods~\cite{Blankenbecler:1981jt,White:1989zz}). We have
\beq
\det D_{\sigma} = \det B^{N_t} \det(\mathbf{1} + e^{N_t\hat\mu_\sigma}
B^{-1}C(A_{N_t-1})...B^{-1}C(A_{0})) \,.
\label{eq:hubb}
\eeq 
The cost of calculating the fermion determinant is reduced from a complexity of ${\cal O}((N_t N_x)^3)$ to ${\cal O}(N_t N_x^3)$. Further simplifications appear because the matrices $C(A)$ are
diagonal, so their multiplications has only a cost linear in $N_x$ and because matrix $B$ does
not depend on the fields $A$ and its determinant and inverse are calculated only once. For periodic
boundary conditions the eigenvectors of $B$ are plane waves 
$(\psi_k)_x=\exp(i k x)$ with $k$ an integral multiple of $2\pi/N_x$. The inverse
is then
\beq
(B^{-1})_{xx'} = \frac1{N_x}\sum_{k} \frac1{\lambda_k} \cos{k(x-x')} \,, \quad\text{with}\quad
\lambda_k = 1 + 2\hat\gamma \sin^2 (k_x/2) \,.
\eeq

\begin{figure}[t]
\includegraphics[width=0.45\textwidth]{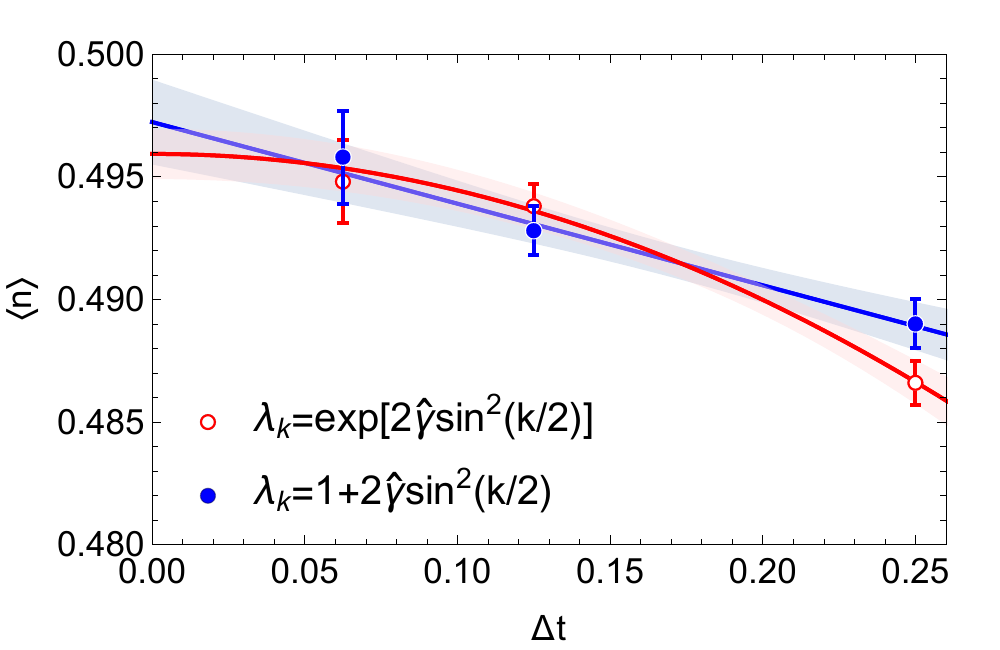}
\includegraphics[width=0.45\textwidth]{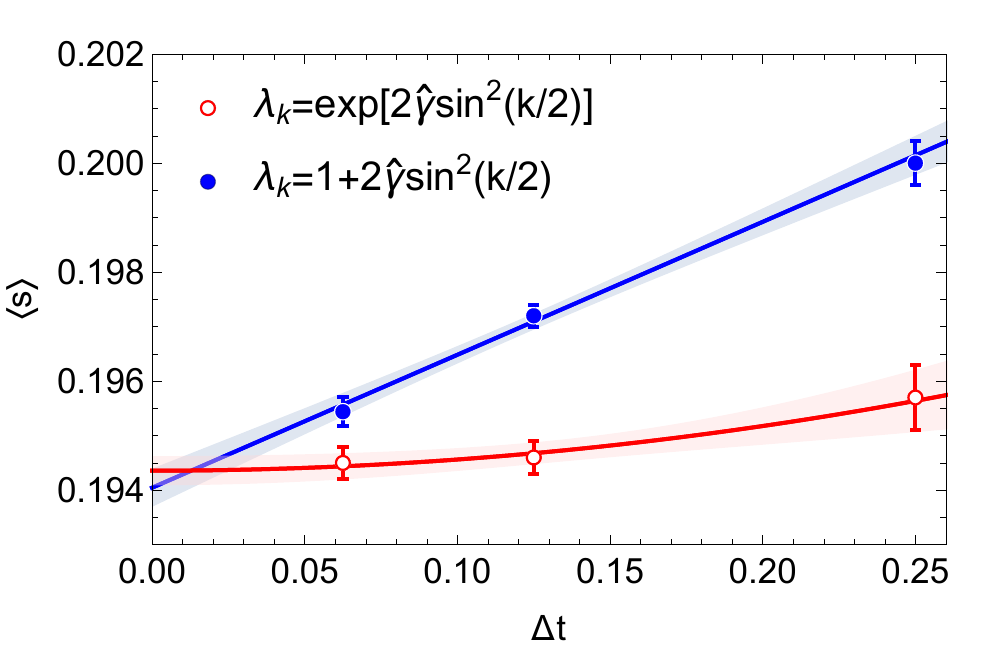}
\caption{Continuum time limit for the density (left) and spin (right) at $\beta \mu = 2.0$ and $\beta h =2.0$. The spatial lattice spacing is fixed to $\Delta x=1$ and the number of sites to $N_x=61$. The extrapolation is done with a quadratic correction for the exponentiated form and with a linear correction for the other case.}
\label{fig:cont-dt-lim}
\end{figure}

As it turns out the partition function in the reduced form as it appears in 
Eq.~\ref{eq:hubb} is very similar to the one for the Hubbard model~\cite{White:1989zz}: 
the fermionic contribution in given by a similar product of matrices with a diagonal 
matrix including the auxiliary field contribution and a non-diagonal matrix, similar to $B$, 
that encodes spatial hopping. The main difference is that the off-diagonal matrix
is $\exp(\Delta t K)$, where $K_{xx'}\propto \delta_{x,x'+1}+\delta_{x,x'-1}$ is the spatial
hopping matrix, whereas for us $B\propto 1+\Delta t K$. 
The Trotter time discretization used for the Hubbard model has errors of the order 
${\cal O}(\Delta t^2)$~\cite{White:1989zz,PhysRevB.33.6271},
whereas our discretization has errors ${\cal O}(\Delta t)$ (this can be seen by considering that
the two discretizations differ at order $\Delta t^2$ for one step, but the full expression 
requires $\beta/\Delta t$ steps.) A simple way to improve our calculation is to replace
the $B$ matrix above with the exponentiated expression. This amounts simply to replacing
the eigenvalues in the precalculation of the inverse of the $B$ matrix with 
$\lambda_k=\exp[2\hat\gamma \sin^2(k/2)]$ rather than $1+ 2\hat\gamma \sin^2(k/2)$.
In Fig.~\ref{fig:cont-dt-lim} we show the continuum time limit using the two versions of the $B$
matrix, for two observables used in this study. We see that they both converge to the same limit,
but the exponentiated version of the $B$ matrix converges faster, as expected.

\begin{figure}[t]
\includegraphics[width=0.45\textwidth]{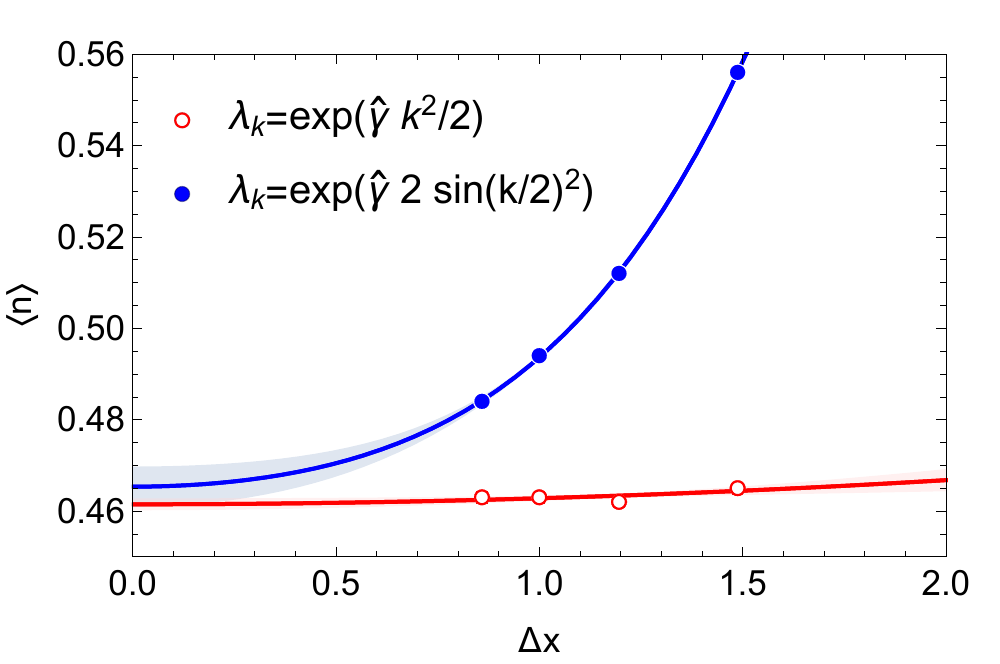}
\includegraphics[width=0.45\textwidth]{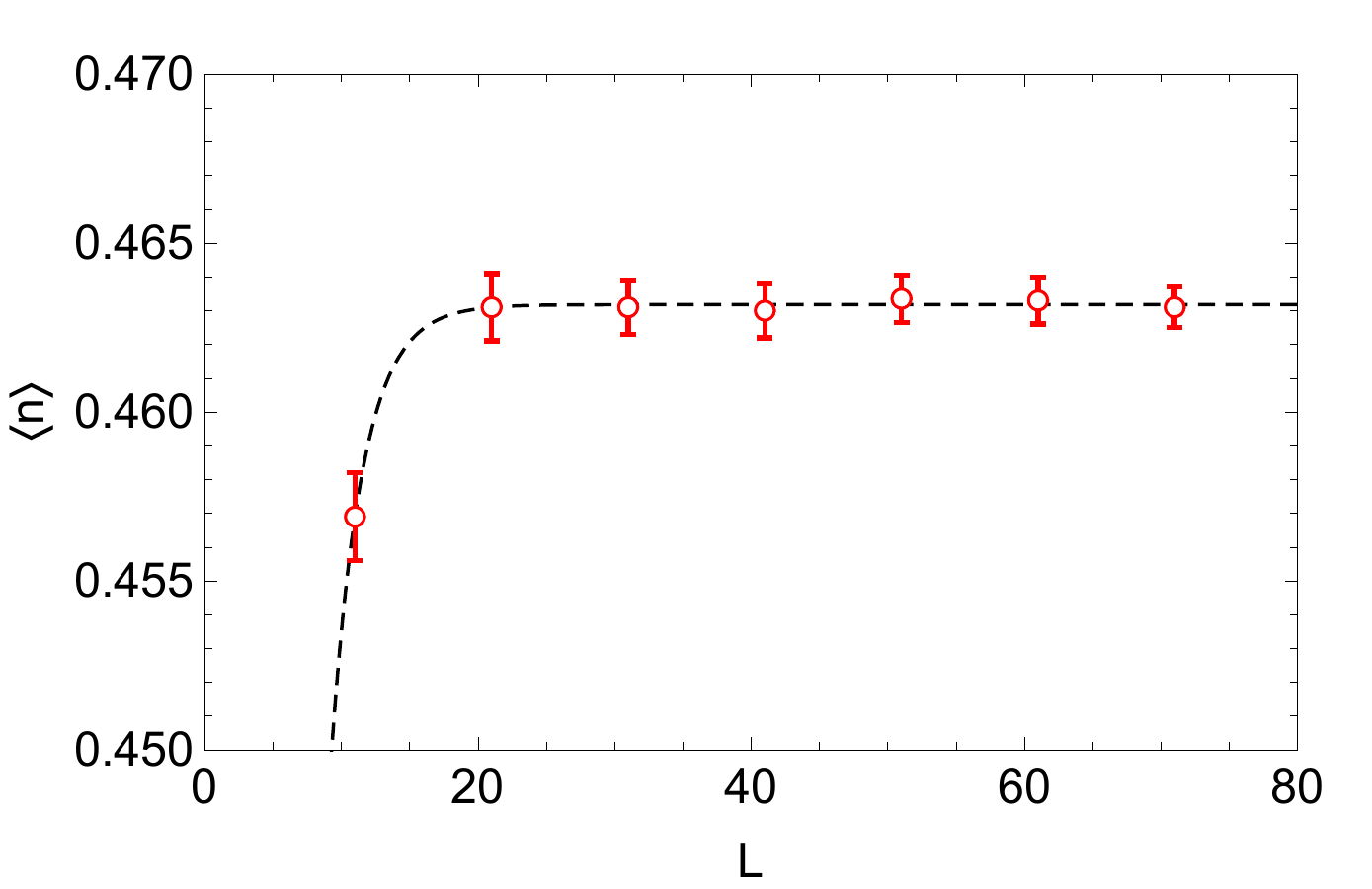}
\caption{Left: Continuum limit in space for the density at $\beta \mu = 2.0$ and $\beta h =2.0$. The spatial lattice size is fixed to $L=61$. The extrapolation is done with a quadratic correction for the continuum dispersion form and with a quadratic plus quartic correction for the other case. Right: Thermodynamic limit for the density for $\beta \mu = 2.0$ and $\beta h =2.0$.}
\label{fig:cont-dx-thermo-lim}
\end{figure}

In a similar vein, we can improve the behavior of the space continuum limit. Note that
using the exponentiated form the eigenvalues of the $B$ matrix are 
$\lambda_k=\exp[2\hat\gamma \sin^2(k/2)]=\exp[\Delta t \hat E(k)]$ with
$\hat E(k)$ the lattice dispersion relations which for small $k$ approximate the
continuum ones $\hat E(k)\approx E(k) = (k/\Delta x)^2/2$. A simple improvement
is to replace the dispersion relations in the eigenvalues with the continuum form,
that is use $\lambda_k=\exp[\gamma k^2/2]$ with $k\in\{-\lfloor N_x/2\rfloor,
 \ldots,-1,0,1,\ldots, \lfloor N_x/2\rfloor \}$
in units of $2\pi/N_x$. In the left panel of Fig.~\ref{fig:cont-dx-thermo-lim} we show the 
continuum limit
using the lattice and continuum dispersion relations. We can see that both converge to
the same limit but the convergence is much faster for the calculation using the
continuum dispersion relation. In the right panel of Fig.~\ref{fig:cont-dx-thermo-lim} we
look at the thermodynamic limit. For this calculation we use the exponentiated form for the
$B$ matrix and the continuum dispersion relations. We set $\Delta t=0.025$ and
$\Delta x=1.0$ which produce values indistinguishable from the continuum limit at the
level of stochastic error bars. We see that for this rather high density, the thermodynamic
limit can be achieved with $L\geq 20$. 
 
Our simulations were carried out using two sampling methods: a straightforward Metropolis 
sampling and Hybrid Molecular Dynamics (HMC). In the Metropolis method we generated new fields by
proposing {\em global} random changes in the auxiliary field $A$, with the shift at every point
limited by a step-size parameter adjusted to produce an acceptance rate around 50\%.
The probability distribution used in the accept/reject step is the one given in 
Eq.~\ref{eq:prob-dist}, with $\det D$ as given by Eq.~\ref{eq:hubb} but with the
matrix $B$ replaced with $\tilde B$ associated with the continuum dispersion relations
\beq
{\tilde B}_{xx'} = \frac1{N_x}\sum_{k} {\lambda_k} \cos{k(x-x')} \,, \quad\text{with}\quad
\lambda_k=\exp[\gamma k^2/2] \,.
\eeq
For the HMC we have to evolve the field $A$ and its conjugate momentum $\pi$ according
to the canonical equations of motion induced by the Hamiltonian $H=\sum_{xt} \pi_{xt}^2/2+S(A)$.
The force term is derived from the integrand in Eq.~\ref{eq:prob-dist}
\beq
\dot\pi_{xt} = -\pd{S}{A_{xt}} =  -\frac1{\hat g}\sinh A_{xt} +
 \sum_\sigma [1-(1+e^{N_t \hat\mu_\sigma} U_t(A))^{-1}]_{x;x} \,,
\eeq
where $U_t(A) = {\tilde B}^{-1}C(A_{t+N_t-1})...{\tilde B}^{-1}C(A_{t})$. The HMC sampling
requires the determinant to be positive, but this seems to be the case in all our simulations;
we will discuss this point below in more detail. We carried out simulations for many different
parameters using both Metropolis and HMC sampling and the results agreed in all cases.

A point to note is that in the evaluation of matrix $U_t(A)$  for large $N_t$, required for both Metropolis and HMC
sampling, the product matrix $U$ exhibits numerical instabilities. This problem is
encountered in Hubbard model simulations and it was found that the instability can be 
overcome by using a factorization of the intermediate results~\cite{White:1989zz}:
the product is split into subproducts of matrices that can be computed directly intermixed
with factorizations. A similar instability appears when computing the force term for
the HMC algorithm: A simple optimization is to compute $[1+\exp(N_t \hat\mu_\sigma)U_0(A)]^{-1}$ 
and then evaluate the other time-slices using the property 
\beq
[1+e^{N_t \hat\mu_\sigma}U_t(A)]^{-1} = 
[C(A_{t-1})^{-1}\tilde B]^{-1}[1+e^{N_t \hat\mu_\sigma}U_{t-1}(A)]^{-1}
[C(A_{t-1})^{-1}\tilde B] \,.
\eeq
The matrices $C(A_t)$ are diagonal and trivial to invert and the $\tilde B$ matrix 
and its inverse are precomputed. This iteration can be used safely for a small number of 
time-slices, but for large $N_t$ we need to recompute $U_t(A)$ from time to time. In our runs
we found that subproducts of up to 20 matrices can be performed safely when using double
precision arithmetic.

\begin{figure}[t]
\includegraphics[width=0.47\textwidth]{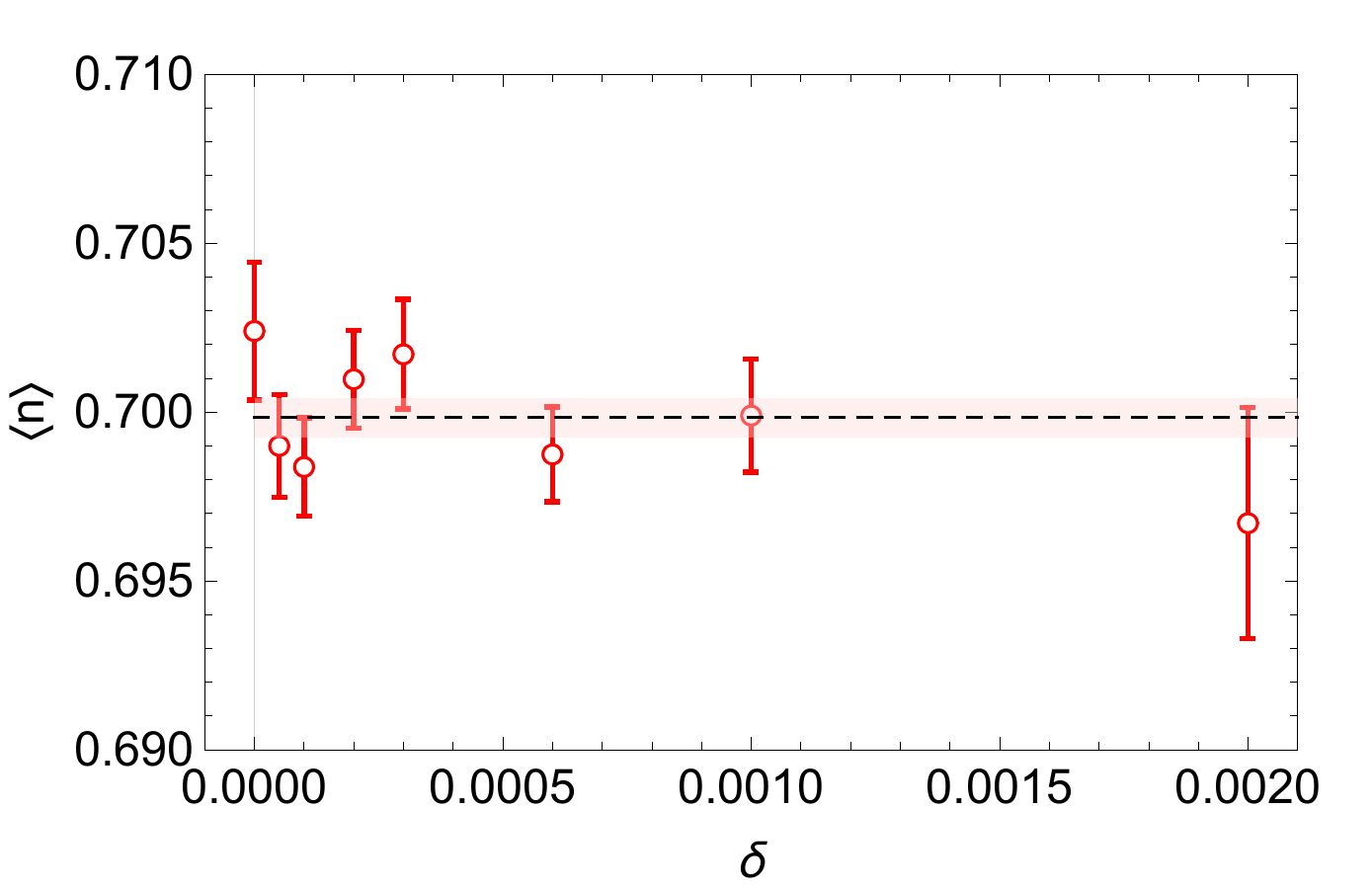}
\includegraphics[width=0.45\textwidth]{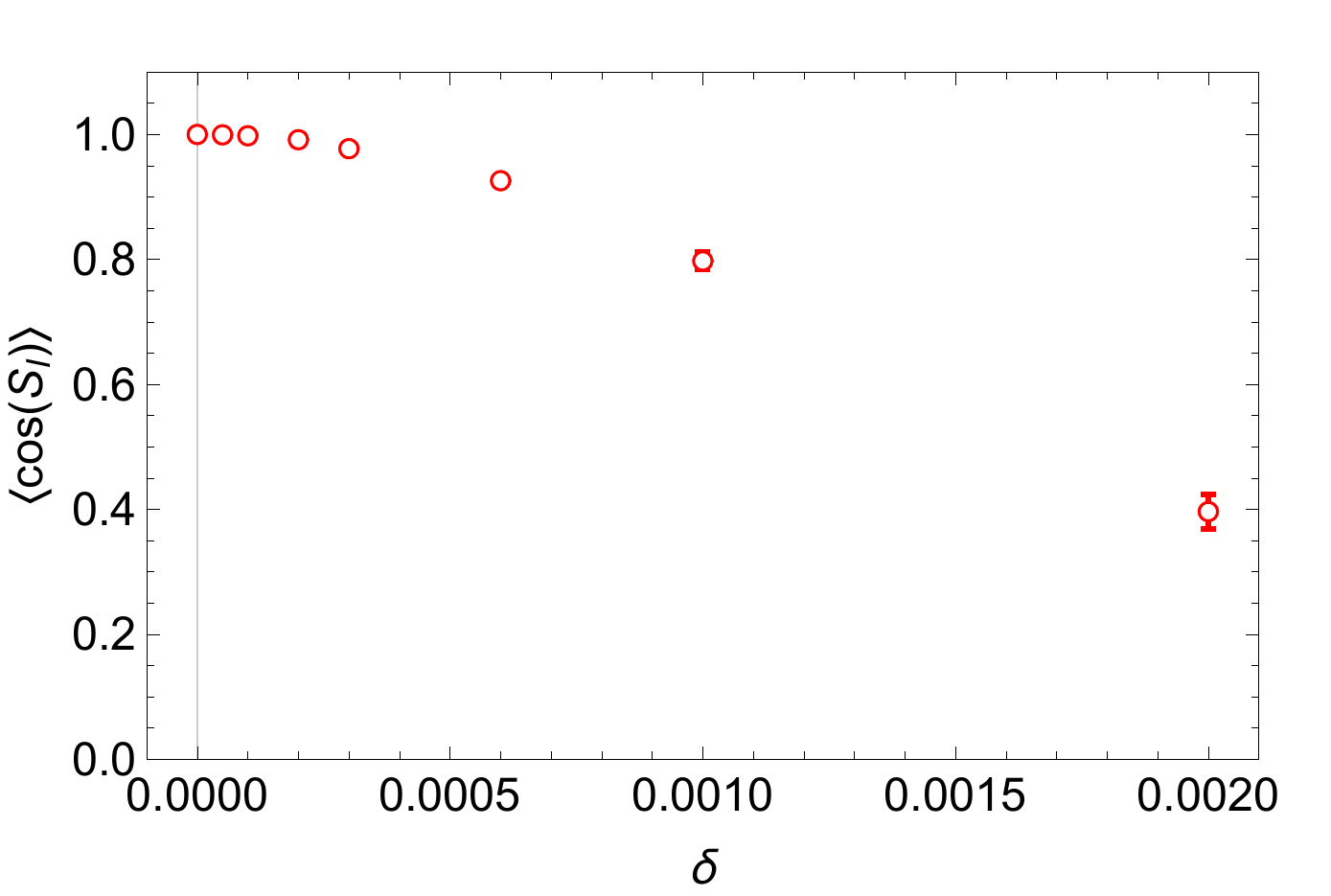}
\caption{Results for the shifted integration contour $A_{xt} \rightarrow  A_{xt}+i\delta$ 
as a function of $\delta$. On the left we have the density and in the right panel we plot the value of the phase average $\av{\cos(\Im S)}$. The horizontal line in the density plot corresponds
to a common fit to all data points.
}
\label{fig:planeshift}
\end{figure}

One final issue to discuss for our numerical simulations is the possibility of trapping
in the positive determinant region of the configuration space. This issue arises even
for non-polarized simulations where $\det D = (\det D_\up)^2$ and the determinant is positive.
While the determinant is positive, the individual spin determinant $\det D_\up$ changes
sign and the positive and negative regions are separated by borders where the determinant
is zero. Since both Metropolis and HMC sampling rely on small (or smooth) changes in 
the configuration, the edges of the same sign regions act as potential barriers and it
is possible that the simulation becomes trapped. This lack of ergodicity was studied
in the context of Hubbard model for both Metropolis sampling~\cite{PhysRevB.44.10502}
and HMC~\cite{PhysRevB.97.085144}. For polarized systems the same-sign regions of
$\det D_\up$ and $\det D_\down$ are not identical: as the polarization increases 
the borders between these regions 
are broadened and the determinant $\det D$ has fluctuating signs. Since we do not see
any sign fluctuation, we believe that the probability distribution we sample has small
(almost vanishing) overlap with these regions. 

As a further test, we changed the integration
manifold to avoid trapping, a method inspired by our work on thimbles~\cite{Alexandru:2015sua}.
The basic idea is that a generalization of the Cauchy theorem guarantees that the path
integral remains unchanged for rather large set of deformations of the integration domain 
in the complex field space. We consider here a shift of the integration domain in the
imaginary direction, which amount to adding a constant imaginary component to each field
variable: $A_{xt}\to A_{xt}+i\delta$. While the integral remains the same, the integrand
becomes complex and exhibits phase fluctuations. For small shifts,
the fluctuations will be mild except close to the regions where the determinant changes
sign. As we move away from the real integration domain the potential barriers associated
with the zero determinant  are lowered, removing trapping, but the sign fluctuations 
become larger. If the results are biased by trapping we should see a change in the value
of the observables as we vary the imaginary shift $\delta$ away from zero.
In Fig.~\ref{fig:planeshift} we plot the density and the average phase as a function of $\delta$
for one ensemble. The parameters used for this ensemble are $G=1/\sqrt 8$, $L=61$ as for the
other tests, but we move to lower temperature $\beta=8$ and higher densities and polarizations
$\beta\mu=4$ and $\beta h=4$ as trapping is expected to appear more readily at low temperature
and high densities. The lattice parameters are set to $\Delta x=1$ and $\Delta t=1/20$. We
see that the density remains unchanged as we vary $\delta$ while the sign fluctuations become
significant. The spin density is also unaffected by the shift. 
We conclude that it is unlikely that our simulations are trapped.

 
\section{Results}\label{results}

\comment{
 Throughout our simulations, we set the scales by choosing the spatial lattice spacing $\Delta x = 1$. Before detailing the energy scales, recall that in non-relativistic natural units every physical quantity can be expressed as a power of length. The uncertaintly principle $\Delta x \Delta p \sim 1$ shows that momentum scales as inverse length $p = [L]^{-1}$ and Schr\"odinger's equation shows that energies scale as $E = [L]^{-2}$. With these facts in mind, we set the inverse temperature $\beta/\Delta x^2= 8$ and the dimensionless coupling $\lambda \equiv \sqrt{\beta} G = 1.0$. All of our simulations are carried out on a lattice with a fixed number of spatial lattice sites $N_x = 61$ and in each simulation 500 measurements are taken. For each set of physical parameters, a time continuum limit is taken. At fixed $\beta/\Delta x^2 = 8.0$, we take a sequence of $N_t = 32, 64, 128$ time slices and extrapolate to the continuum time limit. We find that a linear extrapolation is sufficient to describe the data. A spatial continuum limit is not taken, however this is not much of an approximation as we maintain our systems dilute enough that the wave-numbers of particles populating the system are long enough that the spatial lattice cannot be resolved.
}

The observables that we focus on here are the number and spin density: 
\beq
\av n = \av{n_\up+n_\down}=\frac{1}{\beta L} \av{ D^{-1} \frac{\partial D}{\partial \hat \mu}} \quad\text{and}\quad
\av s = \av{n_\up-n_\down} = \frac{1}{\beta L} \av{ D^{-1} \frac{\partial D}{\partial \hat h}} \,,
\eeq
where the average is taken over Monte Carlo configurations. 
For our simulations we set $G=1/\sqrt 8$ and $\beta=8$. The results presented are computed
using $\Delta x=1$ and $\Delta t=1/8$ which are indistinguishable from the continuum results
at the level of the error-bars (see Fig.~\ref{fig:cont-dt-lim} and 
Fig.~\ref{fig:cont-dx-thermo-lim}). The volume is set to $L=61$ which is close to the
thermodynamic limit. We set the parameters to these values to compare our results with
those from imaginary polarization studies~\cite{PhysRevA.92.063609} and complex Langevin~\cite{Drut:2017fsv}.
All of our results are computed using 500 statistically independent measurements.

\comment{
\begin{figure}[t]
\includegraphics[scale=0.6]{typical-obs.png}
\caption{Continuum limit trajectory of the density (left) and spin (right) at $\beta \mu = 2.0$ and $\beta h =2.0$. Such linear behavior is representative of the densities and polarizations studied in this work. }
\label{fig:typical-cont-lim}
\end{figure} 
}

\begin{figure}[t]
\includegraphics[width=0.45\textwidth]{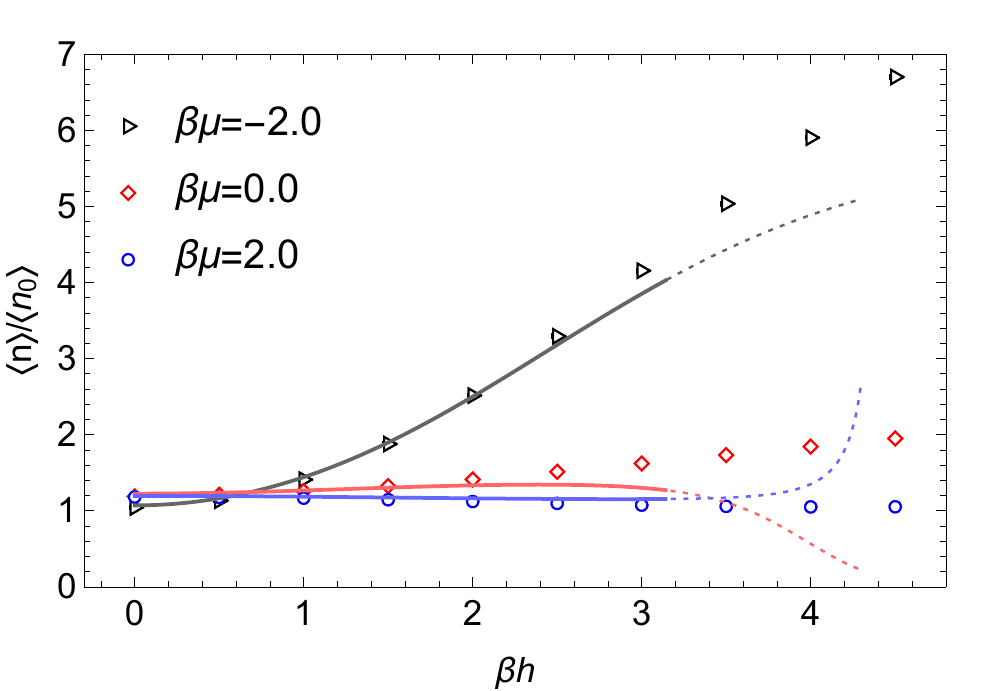}
\includegraphics[width=0.45\textwidth]{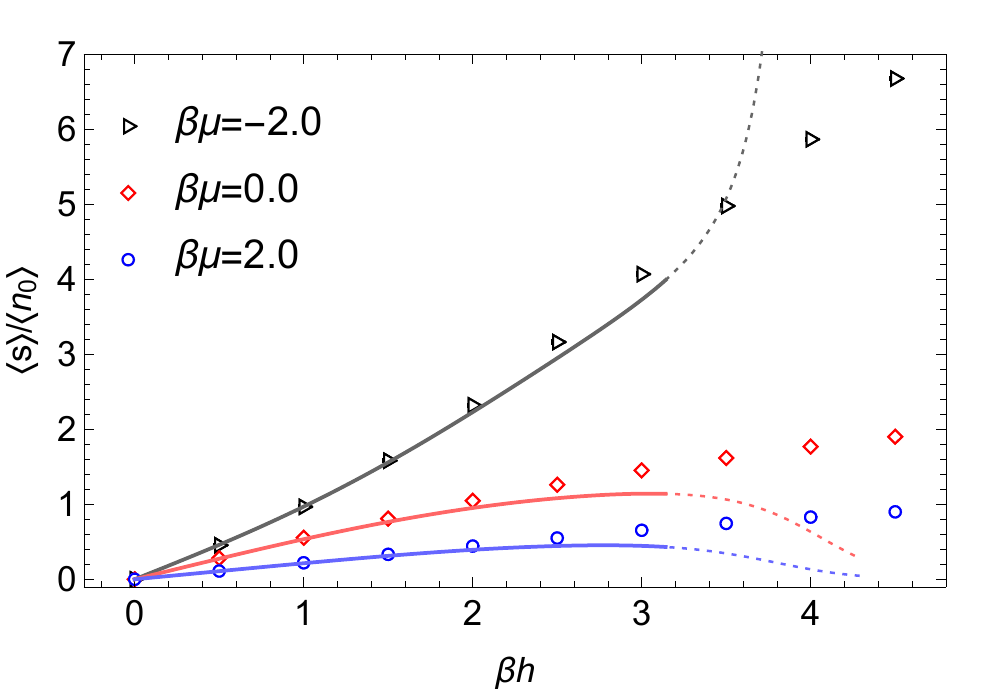}
\caption{Density (left) and spin (right) normalized by the non-polarized density of the free fermi gas $n_0(\beta,\mu)$. The continuous lines show the results obtained in~\cite{PhysRevA.92.063609} by an analytic continuation from imaginary spin chemical potential. This method is expected to work well for $\beta h\leq \pi$, the region indicated by solid lines, but it quickly becomes unreliable as we move to higher polarizations, as we can see by
comparing our results with the dotted lines.
}
\label{fig:all-obs}
\end{figure}

Our main result is presented in \fig{fig:all-obs}: for three values of $\beta \mu$ at $-2.0$, $0.0$, $2.0$ we sweep the spin chemical potential between $0\leq \beta h \leq 4.5$. In addition to our data, we also include in \fig{fig:all-obs} the result of analytically continuing the results of calculations performed at imaginary chemical potential, a study carried out in~\cite{PhysRevA.92.063609}. One finds at small polarization reasonable agreement between our method and analytic continuation, however at large polarization there is clear disagreement. The imaginary
polarization method is expected to fail for large polarizations, when $\beta h\gtrsim \pi$,
but we see large deviations for polarizations as small as $\beta h\gtrsim 2$. This may
invalidate the expectations that interesting features of the phase diagram lie in the
region where analitycal continuation is reliable~\cite{Braun:2012ww}.

\begin{figure}[t]
\includegraphics[width=0.45\textwidth]{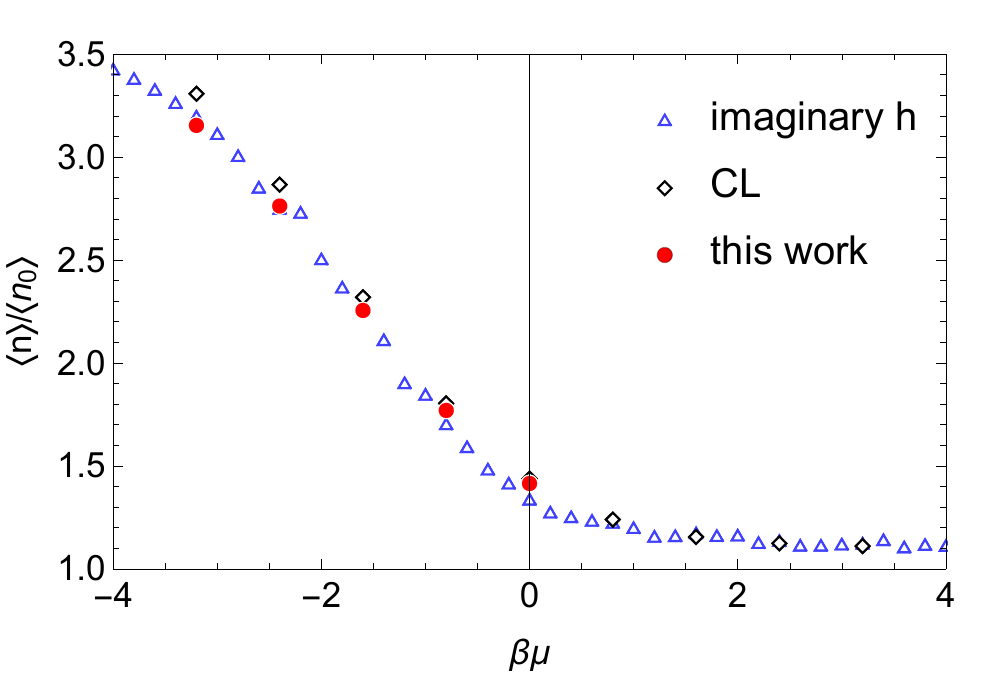}
\includegraphics[width=0.45\textwidth]{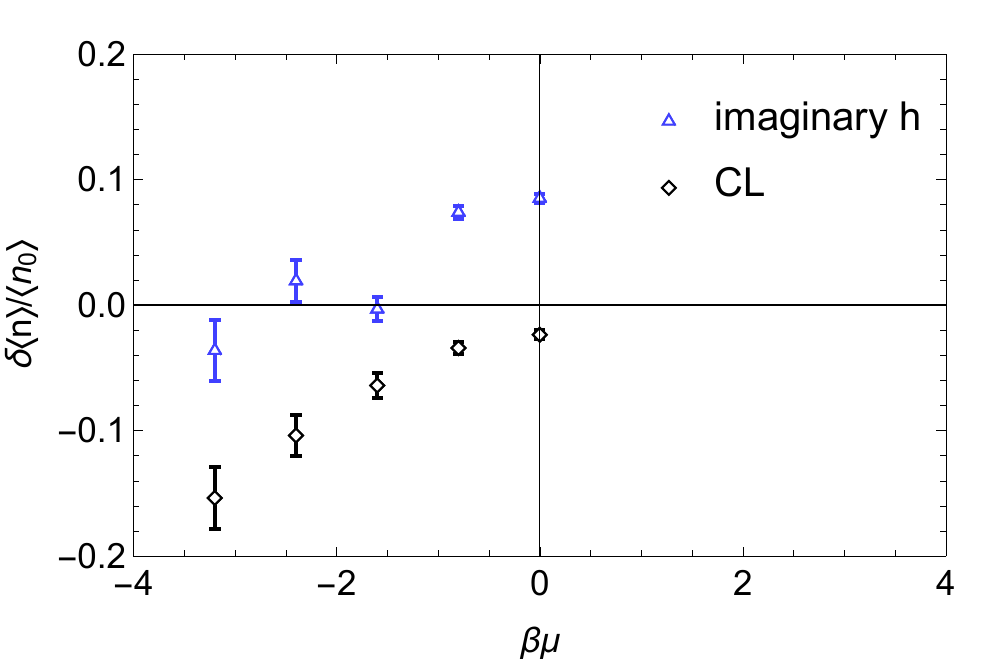}
\caption{Our results for the density compared with results from analytical continuation 
from imaginary polarization~\cite{PhysRevA.92.063609} and complex Langevin 
simulations~\cite{Drut:2017fsv}. 
The parameters for these simulations are $G=1/\sqrt 8$, $\beta=8$, and $\beta h=2.0$.
The right panel indicates the
difference between our results and the other methods.
The error-bars indicate the statistical uncertainties of our results since we did not have information
about the error bars on the results from the other approaches.
}
\label{fig:compare}
\end{figure}

In \fig{fig:compare} we compared our results with analytical continuation and the complex Langevin methods~\cite{Drut:2017fsv}. We use a polarization of $\beta h=2$ and focus on the low chemical
potential region where there seems to be a tension between the analytical continuation and
the complex Langevin results. In this region our results seem to agree at low density with
the imaginary-polarization method but as we increase the density the results seem to agree
better with complex Langevin results. Since we do not have error estimates for the other
methods, we cannot make a sharp statement.


\section{Conclusions}\label{conclusions}

We study using Monte Carlo methods the physics of a system composed of  two species of fermions in one spatial dimension, including unbalanced systems with unequal densities of the two species where a sign problem exists. We find that the sign problem is extremely mild for a vast range of parameters corresponding to densities small compared to the lattice cutoff scale which is the relevant parameters region for the continuum limit. This is in contrast to the parameter values corresponding to densities close to half-filling (but not exactly equal to it), relevant to the physics of the Hubbard model, where a severe sign problem occurs. The simulations in this work were carried out at $G =1/\sqrt{8}$, but we also find no negative determinants in simulations with couplings as high as $G=4/\sqrt{8}$ and it's possible that even higher couplings have mild sign problems.

%

The range of density and spin studied in this work is wide. 
For some parameter values we are able to compare our results to previous calculations using the complex Langevin or the imaginary chemical potential methods. Detailed comparisons are not possible as the published results lack proper error bars but the general lesson is that out calculations are very close to the complex Langevin results (lending credence to its convergence to the right result) and agree with the imaginary chemical potential results at small enough asymmetries. For larger asymmetries  our results are very different from the imaginary chemical potential ones. This is hardly surprising giving the nature of the analytic continuation from imaginary to real chemical potentials.

One potential problem with our calculations is the possibility that the Monte Carlo chain is trapped in a region of positive determinants. Besides producing the wrong result this would give the false impression that the sign problem is milder that it is in reality. Many pieces of evidence were offered to argue against that. The first is that calculations done with the Hybrid Monte Carlo method (more prone to trapping) agrees with the one done with the Metropolis algorithm (which is not likely to be trapped). We also used ideas from our work on the ``thimble" approach to perform a calculation of the path integral not over the real variables, but over a hyperplane slightly shifted on the imaginary direction. Since now the determinants are complex the regions with opposite signs of the determinant are not separated by a repulsive barrier where the determinant vanishes. We observed no difference in the results. Finally, the agreement with other methods for parameter values where these methods are more reliable lend further confidence in our calculation.

It is unclear at this point whether similar methods can be easily extended to higher dimensional systems. Work in this direction is being pursued and results will appear elsewhere.

%
%
%
%

\section{Acknowledgments}
We gratefully acknowledge Joaqu\'{i}n E. Drut and Andrew C. Loheac for helpful comments on the manuscript and discussions. A.A. is supported in part by the National Science Foundation CAREER grant PHY-1151648 and by U.S. DOE Grant No.DE-FG02-95ER40907.  P.B. and N.C.W. are supported by U.S. Department of Energy under Contract No. DE-FG02-93ER-40762. A.A. gratefully acknowledges the hospitality of the Physics Department at the University of Maryland where part of this work was carried out. P.B. and N.C.W gratefully acknowledge the hospitality of the Physics Department of The George Washington University where part of this work was carried out.

\bibliography{non-rel-bib}
\end{document}